# CHALLENGES IN TRANSFORMING, ENGAGING AND IMPROVING M-LEARNING IN HIGHER EDUCATIONAL INSTITUTIONS: OMAN PERSPECTIVE

Dr. Ramkumar Lakshminarayanan *[1]
Mr. Rajasekar Ramalingam *[2]
Mr. Shimaz Khan Shaik *[3]

[1][ramkumar.sur@cas.edu.om]  [2][rajasekar.sur@cas.edu.om]  [3][shimaz.sur@cas.edu.om]

*  *Faculty, Department of Information technology, College of Applied Sciences – Sur Oman.*

**Abstract**

*Nowadays, the student community is growing up with mobile devices and it has becomes an integral part of their life. Devices such as smartphones, tablets, and e-book readers connect users to access information and enabling instant communication with others. The enormous growth and affordability of mobile devices influenced their learning practices. Mobile technologies are playing a significant role in students' academic activities. The factors like convenience, flexibility, engagement, interactivity and easy-to-use enable mobile learning more attractive to students. With these trends in mind, it is important for the educators to inherit the mobile technologies in effective teaching and learning. Our study explores the challenges that exist in implementing the m-learning technologies in the teaching and learning practices of higher educational institutions of Oman. Our study also addressed various issue like adoption of technology, transition to new technology and issues related to engaging students. Based on the outcomes of the study, a framework has been formulated to address all the challenges that are identified for the successful implementation of m-learning.*

**1. Introduction**

The fundamental belief of m-learning is not to convert all computer based learning into a mobile format, but to consider how the mobile devices can be used to strengthen the overall learning strategy [1]. M-learning is learning across multiple contexts, through social and content interactions using the personal electronic devices [2], a sort of learning that happens when the learner is not at a fixed predetermined location or learning that happens when the learner takes advantage of the learning opportunities offered by mobile technologies [3]. With respect to technologies, 'mobile' generally means portable and personal, like a mobile phone [4]. Personal digital assistants (PDA) and mobile phones are the most commonly used technologies for mobile learning which can be broadly categorized on the two dimensions of personal vs shared and portable vs static [5]. The development of m-learning is not intended to replace the classroom learning but enhance the value of Wi-Fi network [1]. Immense growth in information and communication technologies (ICT), fast changing learner behavior, new advancements in the mobile broadband, smart phones, tablets, mobile applications, availability of low cost mobile, wireless devices and infrastructure provides new opportunities for the educational institutions, learners and educators and compels to adopt new technologies in their approaches to pedagogy. The factors influencing the success of m-learning are availability of technology, institutional support, connectivity, integration into everyday life and ownership by the learners [3]. The other factors which attract the learners towards m-learning are convenience, flexibility, engagement, interactivity and easy-to-use attracts learners towards m-learning.  Advantages of m-learning are more flexible, accessible and personalized learning, these advantages hop to keep the learners engaged in ongoing learning activities and enhance their productivities and effectiveness [1]. M-learning potentially brings a reward of placing institutions at the forefront of pedagogical practice and address learners requirements for flexibility and ubiquity [6].

In this paper, a survey was conducted across the colleges of higher education, Oman to identify the most common mobile operating system and the type of internet connection used. Based on the result of study, a framework for implementing m-learning in college of applied sciences (CAS) has been proposed to take full advantage of mobile experience to implement across various programme in CAS, Oman.  An adoption model has been proposed for effective implementation of the proposed framework.  A detailed development methodology is also proposed with the expected challenges to be faced while implementing the proposed model.

## 2. Internet usage and Telecom market in the Sultanate of Oman

Internet users in the Sultanate of Oman have been continuously increasing in recent days. According to the world internet usage statistics news [7], Oman constitutes 2.1% of the worldwide internet users and it has registered an internet usage growth of 66.4% of penetration with 2,139,540 of internet users as on December 31, 2013.  By the mid of 2014, Oman's fixed broadband subscription base (ADSL, WiMAX and leased lines) reached 168,498 translating to 97.8% of total fixed Internet accounts of earlier dialup option [8].

| Market Segment | 2009 | 2010 | 2011 | 2012 | 2013 |
|---|---|---|---|---|---|
| Subscribers: | | | | | |
| Mobile Subscribers | 3,970,563 | 4,606,133 | 4,809,248 | 5,277,591 | 5,617,426 |
| Fixed Telephone Lines | 300,139 | 283,941 | 287,323 | 304,545 | 351,411 |
| Fixed Internet | 78,135 | 73,908 | 89,063 | 119,398 | 158,678 |
| Fixed Broadband | 41,114 | 52,630 | 78,217 | 113,324 | 154,290 |
| Active Mobile Broadband | 425,398 | 734,091 | 1,076,254 | 1,646,098 | 2,443,296 |
| Penetration: | | | | | |
| Mobile Penetration | 125.1% | 166.08% | 173.4% | 160.16% | 155.05% |
| Fixed Telephone Line Penetration 100/inhabitant | 9.46% | 10.20% | 10.40% | 9.24% | 9.70% |
| Internet Penetration (subscribers/100 household) | 22.80% | 18.40% | 22.10% | 29.68% | 39.44% |
| Active Mobile Broadband Penetration (subscribers/100 inhabitant) | 15.30% | 26.50% | 38.80% | 50.00% | 67.44% |

**Table 2.1: Market status for five year**
**(Source: TRA Annual Report 2013)**

As per the TRA annual report 2013 [9], the telecom sector has been growing steadily and has shown a modest upward trend during 2013.  The broadband market for fixed and mobile broadband have witnessed substantial growth in the past two years in terms of the number of subscribers; especially with the introduction of 4G/LTE speeds in the mobile broadband market. The mobile subscribers have increased over the five years and continuously to the year 2013.  It has registered 5,617,426 subscribers with an increase of 6.4% as compared to year 2012.  The penetration rate reached to 155%.  Table 2.1 shows the market status for the five years in the

Sultanate of Oman. The prepaid mobile service subscribers accounted for 91.2% of the subscriber's base with 5,121,723 and the remaining of 8.8% being 495,703 postpaid mobile service subscribers. The year 2013, witnessed an increase in the number of internet subscribers by almost 32.9% to reach 158,678 subscribers compared to 119,398 by end of 2012. In the year 2013, the fixed internet penetration rate was 4.4%, while penetration per household achieved was 39.4%. Figure 2.1 shows the mobile penetration rate per 100 inhabitants. At the end of year 2013, it is witnessed that the active mobile broadband subscribers is continuously increasing and reached to 2,443,296, with a penetration rate of 67.4%.

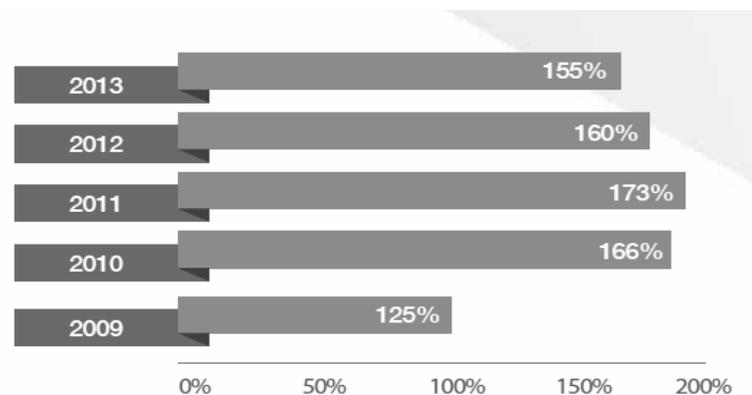

**Figure 2.1: Mobile penetration rate per 100 inhabitants**
**(Source: TRA Annual Report 2013)**

## 3. Literature Review

The best practice for teaching and learning can be achieved by using technology and it is necessary to use online methods as an additional tool in the new context of teaching and learning. Learning that takes place with the help of mobile devices is defined as m-learning [10]. Based on Sharples, use of mobile or wireless devices while in motion for the purpose of learning is m-learning [11]. A balance of technical and human demands by the lectures to achieve m-learning goals and to maintain fidelity with their existing theories about teaching and learning is described as new digital pedagogy [12]. M-learning can be used to encourage both independent and collaborative learning experiences and helps learners to identify areas where they need assistance and support and helps to raise self-esteem and self-confidence [13]. There are other studies which questions the suitability of mobile devices for formal learning as the over-hyping of new devices is compelling to adopt their use in learning [14]. Laurillard [15] recommends that educators need to investigate theirs requirements with the suitability of new technology to accommodate technology in education rather than adapting education to technology. Koszalka and Ntloedibe-Kuswani insisted the m-technology integration and suggested a development of framework for evaluation that supports m-learning program improvement and learning assessment. The information and research findings about effective pedagogical practices in conjunction with mobile technology are in high demand among the educators. There are more research publications in the field of m-learning and it promotes literatures promote various elements that constitutes m-learning in the educational institutions. Still there are arguments whether to design the context for the technology or to adopt technology for learning. There are studies to find the suitability of the mobile devices for personalized learning. Most of the authors quote the benefits and also warns to ensure the stability of education. Some authors use the existing pedagogy framework and others redesign their concepts in new form of teaching and learning. There is a need for professional development for educators to use mobile technology effectively in teaching and learning practices. There are also challenges for students to distinguish between the formal and informal learning. The effects of m-learning practices can be achieved by means of designing a framework specific to the educational domain and to pedagogical practices.

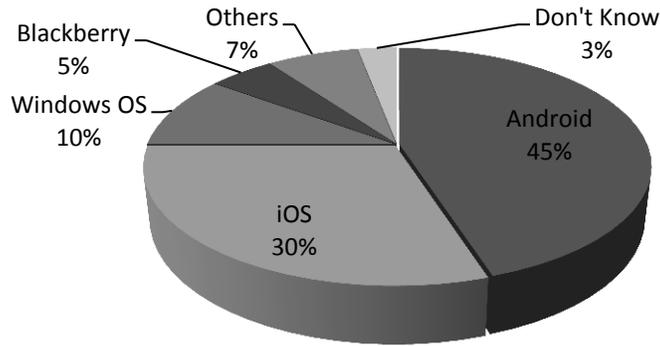

**Figure 4.1: Most common Mobile OS in use (CAS Perspective)**

**4. Proposed model**

Educators believe m-learning offers learner greater access to relevant information, reduced cognitive load and increased access to other people and systems. Plenty of resources available for all the programme of College of Applied Sciences (CAS), Ministry of Higher Education (MoHE), which can be accessed by associated and authentic learner of CAS. A learner can choose any form of resources (web page, audio or video tutorials), send a query to an educator, communicates an expert for practice and guidance using any suitable mobile device.

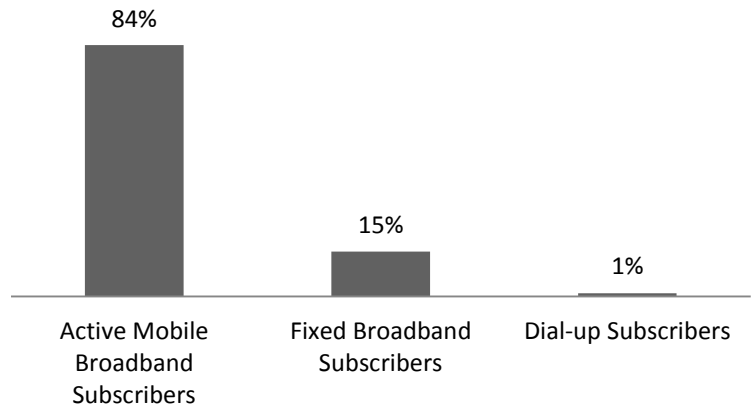

**Figure 4.2: Internet access – Internet connection types used by respondents**

A survey was conducted across the colleges of CAS to identify the top mobile operating system (OS) and the internet connection type used by the educators and learner. The survey reveals, 45% of the respondents use android mobile OS, which is very popular and the top mobile OS used by the educators and the learner of CAS.  Figure 4.1 shows most common mobile OS used by the respondents.  Figure 4.2, shows the various types of internet connection types used by the respondents.  84% of respondents use mobile internet, of these, 40% use Wi-Fi to access the internet and 44% use packet data connection (prepaid internet connection).  Another major finding is that the majority of the learner and educators are using the latest mobile devices for maximum hours.  Through the study, it has been identified that the course support, evaluation of assessment, interaction and communication between the learner and educator can be done effectively and purposefully using mobile devices.  In order to develop the content for the course offered by the CAS, there is a need for a framework, specific to m-learning due to the unavailability of specific framework to incorporate the m-learning in CAS, Oman.

In this work, a framework for m-learning in CAS (FMLCAS) is proposed for effective implementation of m-learning in CAS, Oman, which is an extension of the features of FRAME model proposed by Koole (2006).  Figure 4.3 shows the FMLCAS framework which includes various aspects such as device, learner, educator, system coordinator (SYSCO), programme director (PD) and m-learn administrator (ML-ADMIN).   The framework provides an environment for the related aspects to interact with each other.

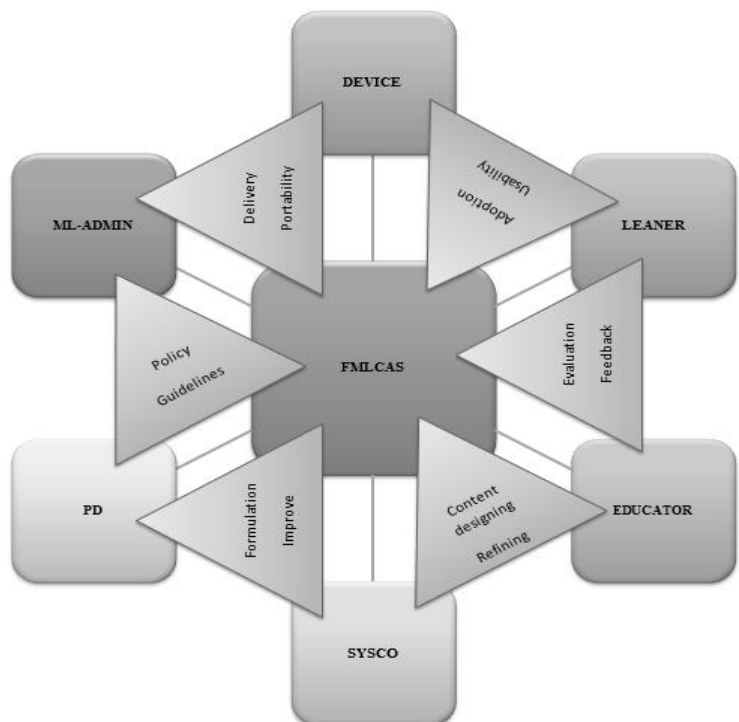

**Figure 4.3 FMLCAS model**

## *4.1 FMLCAS Aspects*
### *4.1.1 Device*
Device refers to the characteristics of mobile devices supporting the requirements of m-learning.  It is important to access these characteristics because it provides the interface between the device and learner.  Learner equipped with latest and high-end mobile devices will be able to focus more on the learning task.  The device criteria refer to the hardware, software, memory storage, I/O and functional capabilities.

### 4.1.2 Learner

This aspect describes how students use their prior knowledge, interpret and transfer information. The criteria to be evaluated are existing expert level, expectation, suitability and willingness.

### 4.1.3 Educator

The educator aspect takes into the accont of the process of identifying the interest and motivation towards m-learning. The criteria to be evaluated are motivation of teaching using mobile devices and adoptability of new technology.

### 4.1.4 Programme Director

This aspect describes m-learning policy and guidelines to be set and to evaluate the curriculum adoptability to m-learning. The criteria to be evaluated are the suitability of the existing courses and the stability of the outcome.

### 4.1.5 System coordinator

This aspect determines what and how information should be learnt and determines the relationship between knowledge production and knowledge utilization specific to m-learning and also supports the process of continually reshaping the course curriculum to the content delivery. Criteria to be evaluated are identifying the pattern of the course and its relationship, relevance and accuracy.

### 4.1.6 M-Learn administrator

This aspect identifies the learner content, authenticity of context and audience and community of practice. ML-ADMIN interacts with the other aspects like device, learner, educator, PD and SYSCO. Each aspect has varying degrees of control over the learning process and varying levels of effectiveness depending on the situation, learner and task.

## 4.2 Intersections
### 4.2.1 Device-Student

This intersection contains elements that belong to both the device and learner aspects. It relates the characteristics of the mobile devices and psychological comfort and satisfaction of the learners. The major criteria to be considered in this intersection are adoption and usability.

### 4.2.2 Learner-Educator

This intersection contains elements that belong to both the learner and educator aspects. It relates the characteristics of the learner aspect, which is the psychological comfort and availability of the educator aspect. The major criteria to be considered in this intersection are evaluation and feedback.

### 4.2.3 Sysco–educator

This intersection contains elements that belong to both the SYSCO and educator aspects. The characteristic of this intersection are providing related content of all courses. The major criteria to be considered in this intersection are context designing and refining.

*4.2.4 Sysco–PD*

This intersection contains elements that belong to both the SYSCO and PD aspects. It relates the characteristics of the SYSCO aspect, which provides the related content for each course of the CAS and PD aspect provides the policies and the guidelines. The major criteria to be considered in this intersection are formulation and improvement.

*4.2.5 M-learn admin–PD*

This intersection contains elements that belong to both the ML-ADMIN and PD aspects. It relates the characteristics of the ML-ADMIN aspect, which designs the course content for all the courses of CAS and PD aspect provides the policies and the guidelines. The major criteria to be considered in this intersection are quality and implementation.

*4.2.6 M-learn admin–device*

This intersection contains elements that belong to both the m-learn admin and the device aspects. It relates the characteristics of the ML-ADMIN and device aspect, which is designing the content suitable for the devices and various properties of the devices in which the m-learn content would be accessed. The major criteria to be considered in this intersection are content delivery and portability.

**4.3 FMLCAS adoption model**

Figure 4.4 shows the proposed model for FMLCAS m-learning adoption, which contains an m-learning environment, policies and guidelines. The proposed model depicts some of the essential elements of m-learning in CAS, including educator, learner, Programme director, M-learn administrator, System coordinators, mobile devices and communication infrastructure.

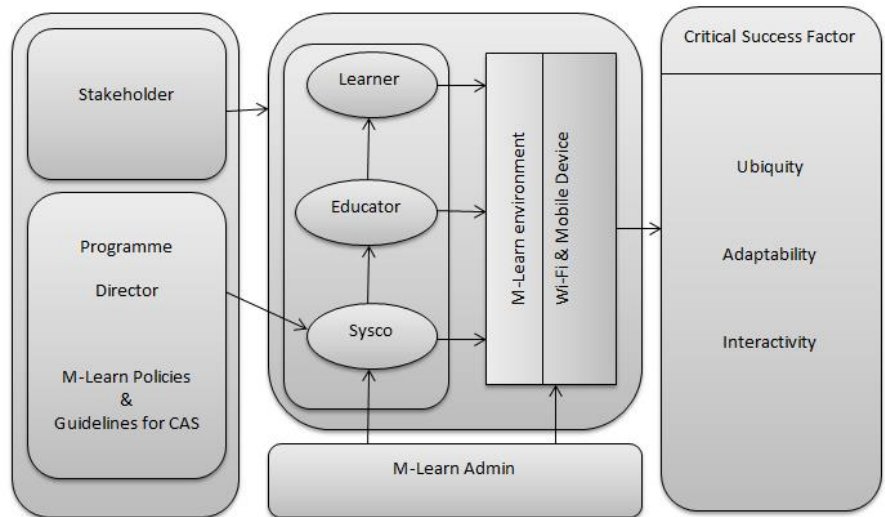

**Figure 4.4 FMLCAS adoption model**

The m-learning environment consists of a communication infrastructure containing Wi-Fi network which enables communication between the mobile devices. Mobile devices used as an academic support for learner and educator in assessment, evaluation, accessing course content, Internet access, learner-to-learner, learner-to-educator and educator-to-educator communication.

When considering the implementation of FMLCAS, CAS must consider their fit within the current curriculum, where a preference should be given to short courses and theory based classes. It is very clear that adoption of m-learning in the CAS context will be influenced by institutional, socio-cultural and intrapersonal factors. Other than this, recommended stakeholders for FMLCAS are college administrators, LRC, support staff, device vendors and parents. Identified stakeholders should be involved in the development of the adoption plans. All these stakeholders and elements have an impact on the m-learning environment either directly or indirectly. CAS need to investigate whether they can provide the training and technical support required for m-learning implementation. The critical success factors are to be evaluated by ubiquity, adoptability and interactivity.

### 4.4 FMLCAS development methodology

A series of development activities must be taken based on the proposed framework to explore and develop new pedagogy for effective m-learning in CAS. To support m-learning within learning environment of CAS, educators must be encouraged and motivated by means of workshop and training programs. Informal sessions and email communication to be used to provide technical updates regularly to all the aspect. Formal and informal education has to be redesigned to adopt the m-learning in part of the learning process. Learner must be encouraged to utilize the benefits of m-learning in all the ways. It recommended including technology experts into the group and discussing the improvement and issues on a regular interval. Activities related to the m-learning are to be framed and a periodical review meeting should be conducted to evaluate the utilization of m-learning. Identify the issues related to the educators through frequent meeting and by collecting the feedback from the learners.

### 4.5 Challenges in m-learning implementation

Learning without an educator managed classroom is possible using m-learning, learner and educators should support with reliable and fast network connectivity to access m-learning resources. In terms of diversion, there is more possibility of using mobile devices for social networking rather than for learner networking. The major consideration and challenge for m-learning environment are the compatibility of the devices and the feasibility of implementing m-learning when devices are heterogeneous. The other challenges include limitations, educator involvement, learner interest, training, safety, security, maintenance and implementation cost. The support team of the CAS, Oman must be expertized in mobile technology and mobile app development to support the m-learning of the educational institutions.

### 4.6 Conclusion

The m-learning FMLCAS framework presented in this paper is an attempt to effectively create an m-learning environment in the teaching and learning practices of CAS. In this paper, it has been identified that there are considerable increase in usage of mobile devices among the learner and educator and also the facility of using the devices using wireless technology is more feasible in Oman. Through literature review the need for the new framework specific to CAS is justified. In FMLCAS, the aspects are identified and an adoption model is framed and recommended. The framework is designed to the suitability to serve the system design process for future implementation.

# REFERENCES

[1] Cobcroft, Rachel S., Towers, Stephen J., Smith, Judith E., & Bruns, Axel , Mobile learning in review: Opportunities and challenges for learners, teachers, and institutions, 2006.

[2] Crompton, H., A historical overview of mobile learning: Toward learner-centered education, Z. L. Berge & L. Y. Muilenburg Editions, handbook of mobile learning (pp. 3–14), 2013.

[3] Sharples, Mike, Mobile learning: research, practice and challenges, Distance Education in China, 3(5) pp. 5–11, 2013.

[4] Crescente et al, Critical issues of m-learning: design models, adoption processes, and future trends, Journal of the Chinese Institute of Industrial Engineers 28 (2): 111–123, March 2011.

[5] Laura Naismith, Peter Lonsdale, Giasemi Vavoula, Mike Sharples, Literature Review in Mobile Technologies and Learning, University of Birmingham, 2003.

[6] Vavoula, G.N., Sharples, M., Challenges in Evaluating Mobile Learning, in Traxler, J., Riordan, B., Dennett, C. (eds) Proceedings of the mLearn 2008 Conference (School of Computing and Information Technology, University of Wolverhampton) pp. 296-303, 2008.

[7] Miniwatts Marketing Group, 2010 Internet World Stats,
Source: http://www.internetworldstats.com/stats.htm).

[8] http://www.arabadvisors.com/reports/item/14898

[9] Telecommunications Regulatory Authority (TRA) Oman, Annual report – 2013.
Source: http://tra.gov.om/annual-reports/1062-annual-reports-2013

[10] Quinn, R. E., Faerman, S. R., Thompson, M. P., & McGrath, M. R., Becoming a Master Manager: A Competency Framework (3 Ed.). New York: John Wiley and Sons, 2003.
.
[11] Sharples, M., Corlett, D. and Westmancott, O, The Design and Implementation of a Mobile Learning Resource. Personal and Ubiquitous Computing, 6, pp. 220-234, 2002.

[12] Lloyd, M., & Irvine, S., Digital pedagogy: Finding the balance in an online learning and teaching environment, proceedings of ASCILITE, 2005.

[13] Attewell, J, Mobile technologies and learning: A technology update and m-learning project summary, London, Learning and Skills Development Agency, 2005.

[14] Melhuish, K., & Falloon, G., Looking to the future: M-learning with the iPad Computers in New Zealand Schools: Learning, Leading, Technology, 22(3), 2010.

[15] Laurillard, D, Pedagogical forms for mobile learning: framing research questions. In N. Pachler (Ed.), Mobile learning: towards a research agenda (153-175). London: WLE Centre, IoE, 2007.